\documentstyle[preprint,aps]{revtex}
\tightenlines
\def\s{\sigma}
\def\vr{\vec r}
\def\ve{\varepsilon}
\def\no{\noindent}

\def\be{\begin{equation}}
\def\ee{\end{equation}}
\def\ba{\begin{array}} 
\def\ea{\end{array}}
\def\bea{\begin{eqnarray}} 
\def\eea{\end{eqnarray}}

\title{\bf {On the relativistic origin of the kink effect in the chain 
of Pb isotopes}}
   
\author{S. Marcos$^a$, L.N. Savushkin$^b$, M. L\'opez-Quelle$^c$, 
R. Niembro$^a$, P. Bernardos$^d$\\ 
$^a$ \small \em Departamento de F\'\i sica Moderna,
\small \em  Universidad de Cantabria,  
E-39005 Santander, Spain\\ 
$^b$ \small \em Department of Physics,\\ 
\small \em St. Petersburg University for Telecommunications, 
191065 St. Petersburg, Russia\\
$^c$ \small \em Departamento de F\'\i sica
Aplicada, \small \em Universidad de
Cantabria,  E-39005 Santander, Spain\\
$^d$
\small \em Departamento de Matem\'atica Aplicada y Ciencias de la Computaci\'on,\\ 
\small \em  Universidad de Cantabria, E-39005 
Santander, Spain\\ 
}

\date{\today}

\begin{document} 
\maketitle 
\begin{abstract} 

We investigate the origin of the 
kink effect (KE) in the relativistic mean field theory by transforming
the single-particle Dirac equation into a Schr\"odinger-like equation. 
It is found that relativistic self-consistent effects as well as 
contributions coming from the $\rho$ meson determine the actual 
structure of the KE.
However, the spin-orbit force generated by the $\rho$ meson
has no significant influence on the KE.
\end{abstract}  

\bigskip
\bigskip


\newpage

The charge radii ($r_c$) of the Pb isotopes have been measured a few years 
ago with a very high precision \cite{otten}, 
their anomalous kink behavior being 
the most essential feature of these measurements. 
The kink effect (KE) means that the experimental data on $r_c$ 
as a function of A in the isotopic chain 
display a change of the slope at N=126 with a gradual addition of 
neutrons. This phenomenon has been a matter of a detailed discussion by 
different theoretical groups during the recent years 
(see Refs. \cite{taji}-\cite{ring} and references therein).

At the first stage, it has been shown that the density-dependent 
Hartree-Fock model with standard parametrizations of the Skyrme (SHF) 
\cite{taji} or Gogny \cite{priv} forces fails to reproduce the empirical charge isotopic 
shifts in the chain of Pb isotopes.

On the other hand, in Refs. \cite{shar93,shar95} 
the anomalous behavior of the charge 
radii of these isotopes has been studied in the relativistic mean field 
approach (RMFA). The Pb data are well described 
by this type of theory.
In Ref. \cite{shar95} 
it is supposed that the success of the RMFA in 
reproducing the KE is achieved due to the weak isovector dependence of the 
spin-orbit force generated by the RMFA.

In Ref. \cite{rein} the problem of isotopic shifts has been investigated 
with exhaustive detail, both in the SHF and RMFA. 
The authors of Ref. \cite{rein} have proposed a new Skyrme functional SkI4 with a more general 
structure of the spin-orbit force than that of the standard Skyrme 
functional.
They have also succeeded to reproduce the KE, but at the 
price of introducing an extra parameter for the two body spin-orbit 
potential.

As the authors of Refs. \cite{shar93,shar95} state, the standard 
RMFA provides an excellent description of the anomalous kink in the 
isotopic shifts about $^{208}Pb$ without using extra fitting parameters.
In the present paper we shall focus our interest 
into the problem of the origin of the KE in the RMFA
paying special attention to the role of the isovector $\rho$ meson in the 
kink structure.

In the RMFA \cite{ring}, the relativistic single-nucleon wave function 
$\psi(\vec r)$, for a nucleon with the rest mass M, satisfies the Dirac equation

\be
[-i\hbar \vec \alpha\cdot\vec \nabla + \beta(M+S(\vec 
r))+V(\vec r)]\psi(\vec r)=E\psi(\vec r),
\ee
where $E=M+\ve$ and $\ve$ is the binding energy eigenvalue while 
$S(\vr)$ and $V(\vec r)$ are the scalar and vector potentials, 
respectively, defined as follows:

\bea
S(\vec r)=g_\sigma\sigma(\vec r),\nonumber
\eea
\bea
V(\vec r)=V_\omega(\vec r)+\tau_0 V_\rho(\vr)+\frac{1+\tau_0}{2}V_c(\vr)=
g_\omega \omega_0(\vr)+\tau_0g_\rho\rho_0(\vr)+e\frac{1+\tau_0}{2}A_0(\vr).
\eea

In Eq. (2) $\s$ is the scalar meson field, $\omega_0$ and $\rho_0$ are 
the time components of the $\omega$ and $\rho$ meson fields, 
respectively, while $V_c$ is the Coulomb potential. 
$g_\sigma$, $g_\omega$ and $g_\rho$ are the coupling constants
associated with the respective mesons.
All meson fields satisfy the corresponding Klein-Gordon equations 
(see \cite{ring}, for example).

Eq. (1) can be reduced to a
Schr\"odinger-like equation \cite{mar} which does not contain the 
first derivative of the wave function
[$\tilde \phi (\vr)$]. It reads:

\be
[-\frac{\hbar^2}{2M}\nabla^2+V_{cent}(\vr)+V_{SO}(\vr)]\tilde \phi (\vr)=
\ve(1+\frac{\ve}{2M})\tilde \phi (\vr),
\ee
\no
where the central potential $V_{cent}$ and the spin-orbit potential 
$V_{SO}$ are given by:
\bea
V_{cent}(\vr)=S+V+\frac{S^2-V^2}{2M}+\ve\frac{V}{M}+
\Delta V_{cent},\\[6mm]
\Delta V_{cent}=\frac{\hbar^2}{2M}[\frac{1}{4}W^2
+\frac{1}{r}W+\frac{1}{2}W'],\nonumber\\[6mm]
W=-\frac{S'-V'}{2M+\ve+S-V};\nonumber\\[6mm]
V_{SO}(\vr)=\frac{{\hbar}^2}{2M}\frac{2}{r}W\vec l \cdot \vec s.
\eea

By writing the Dirac equation in this form, we are able to study separately
the influence of the different terms entering Eq. (4) on the KE 
(we can analyze, for example, if their respective effects are of 
relativistic origin or not).

Our calculations have been carried out for the $Pb$ chain of isotopes
around the $^{208}Pb$ nucleus in the framework of the standard RMFA, 
with a linear model (L) \cite{horo81} and two nonlinear models NL-SH 
\cite{shar93b} and NL3 \cite{lala97} containing 
self-interactions of the scalar field. 
For simplicity, we shall mainly concentrate our discussion 
in the three Pb isotopes $^{206}Pb$, $^{208}Pb$ and $^{210}Pb$.
The most stable configuration of the $^{210}Pb$ isotope in the two 
nonlinear models considered contains two neutrons in the $1i_{11/2}$
states. In the L model, the two configurations
in which two neutrons fill up the 
$1i_{11/2}$ or $2g_{9/2}$ single-particle states are almost degenerate
(we shall refer to these two configurations as IC and GC, respectively). 
We approach the $^{210}Pb$  ground state by the IC in both 
linear and nonlinear models, 
although the BCS formalism would be more appropriate in the first case.
Nevertheless, we expect that our conclusions are also valid for the 
L model, at least, at a qualitative level.

The self-consistent results 
for the charge radii corresponding to the exact models
are summarized in Table I, case $a$, and in Fig. 1 
(the $r_c$ are normalized in this figure to 
the experimental value
of the $^{208}Pb$ nucleus).
The KE appears in the three models considered but 
it is more strongly pronounced in models with smaller K modulus values. 
We notice that the KE is quite small for the GC, in 
spite of the fact that the $2g_{9/2}$ state is less bound than the $1i_{11/2}$ 
state and that the radius of the $2g_{9/2}$ state is larger than the
radius of the $1i_{11/2}$ one. Thus, it is not the radius but 
the different neutron density associated with these two states that
is essential to explain the different KE corresponding to IC and GC.
The KE in the chain of $Pb$ isotopes can be only explained if 
single-particle configurations including
$1i_{11/2}$ states significantly contribute to the total nuclear 
ground state for the isotopes with $A>208$.

To understand how the kink is generated by the model
we have begun from studying the influence of the different mesons on the KE.
We have calculated the r.m.s. radii of different components of 
the single-particle potential: $S$, $V_\omega$ and $V_\rho$ as functions 
of the number A, and we have established that it is only the value of $r_\rho$, i.e., 
the r.m.s. radius of the $\rho$ meson potential $V_\rho$, that shows 
a pronounced kink while the other components do not.
Since the $\rho$ meson has an isovector character,
we can expect that it plays a relevant role in the behavior of protons 
when neutrons are added to the nucleus.
Thus, the result mentioned above 
confirms the relevant role of  the $\rho$ meson in generating the KE.

Let us underline that the $\rho$ meson contribution reveals itself in 
Eq. (4) in different terms of the total potential 
$V_{cent}(\ve)+V_{SO}(\ve)$ through the $V$ component. Since the coupling 
constant $g_\rho$ of the $\rho$ meson is comparatively small, the most 
essential contribution of the $\rho$ meson to the total potential comes 
from the (S+V) and $(S^2-V^2)/2M$ terms, where the $\s$ and $\omega$ 
contributions cancel each other to a certain extent.

To study the influence
of the different components
containing the $\rho$ meson field
on the charge radii of the nuclei in more detail,
it is useful to make the 
following replacements in eqs. (4) and (5) corresponding to
$V_{cent}(\vr)$ and $V_{SO}(\vr)$, respectively :
$g_\rho\to x_1g_\rho$ in the $(S+V)$ term, 
$g_\rho\to x_2g_\rho$ in the $(S^2-V^2)/2M$ term,
$g_\rho\to x_3g_\rho$ in the $\frac{\ve}{M}V$ term,
$g_\rho\to x_4g_\rho$ in the $\Delta V_{cent}$ term, and
$g_\rho\to x_5g_\rho$ in the the $V_{SO}$ potential.
After that, we allow the parameters $x_{1-5}$ to 
vary in a continuous and independent way in the interval [0-1].
We have considered 4 different cases 
corresponding to the 4 following combinations of the $x_i$ parameters:
$a$) $x_{1-5}=1$ (already discussed)
$b$) $x_{1} \ll 1$, $x_{2-5}=1$, 
$c$) $x_{1,3-5}= 1$, $x_2=0$, and 
$d$) $x_{1-5}=0$ (equivalent to take $g_\rho=0$).

The results for these cases are collected in Table I, and can be 
identified by the corresponding letter $a$, $b$, $c$ or $d$ 
in the last column.
We have considered the models L, NL-SH and NL3, and the IC for the ground 
state.
Fig. 2 shows the $r_c$ results for the NL-SH model 
normalized to the $r_c$ of the $^{208}Pb$ nucleus.
The case $a$ contains the total $\rho$ contribution and manifests the KE 
in good agreement with the effect experimentally observed.
In the case $b$, in which the $\rho$ contribution to $S+V$ is almost
eliminated,
the kink has completely changed its structure in relation to the case 
$a$, now the kink seems to be, somehow, the mirror image 
of the kink in case $a$.
This fact makes evident the crucial role the $\rho$ meson plays
in the kink structure through the contribution of the ($S+V$) term.

Table I shows that $r_c$ slightly increases when 
the parameter $x_1$ decreases. 
This fact is related to the decreasing of the proton binding energy
with the decreasing of the $x_1$ value 
(the opposite is true for the neutron energy).
However the qualitative change of the kink structure cannot be 
attributed to this modification of $r_c$ (which has only a small effect 
on the kink).
Notice also that $x_1$ is taken equal to $0.2$ 
and $0.1$ (rather than 0) in the L and NL-SH models, respectively,  
to maintain the $3s_{1/2}$ proton state bound enough.

The effect of the $\rho$ meson on the kink
in the region $208 \le A\le  210$ through the term $S+V$ in the 
potential is quite small for the more stable IC.
However, for the GC, a decreasing of the $x_1$ parameter 
produces a sizable turn at the left side of this region of the kink.

With regard to the case $c$ in Fig. 2, 
the effect displayed is opposite to that 
shown in the case $b$ 
and much smaller in magnitude.
This result is a consequence of the opposite role the 
$\rho$ meson field plays in the terms ($S+V$) and $(S^2-V^2)/2M$ of the 
central potential.
For the GC the effect of the $\rho$ meson on the kink
through these two terms remains opposite to the case $b$,
but comparable in magnitude.

Finally, for the case $d$, i.e., the case without $\rho$ meson contribution, 
the behavior of $r_c$ is intermediate
between that corresponding to the cases $a$ and $b$ as it can be appreciated in 
Table I and Fig. 2.

We have verified that the $\rho$ meson contribution through the $\Delta 
V_{cent}$ and $V_{SO}$ terms has only minor influence on the kink. 
As we have explained above, this behavior is expected because the relative 
contributions of the $\rho$ meson to the ($S-V$) term, entering
$V_{cent}$ and $V_{SO}$, are very small.

To understand better the nature of the KE it would be useful to examine 
the partial contributions $r_{ck}$ of different single-particle 
states "$k$" to the total charge radii $r_c$ of the nuclei under 
consideration $(r_c=\sqrt{r^2_{c1}+...+r^2_{cA}})$. Fig. 3 displays
these values for the six bound states (closest to the Fermi level) calculated 
within the NL-SH model for different nuclei in the region $200\le A\le 214$. 
We see that the total value of $r_c$ for 
$A\ge 208$ is mostly determined by the contribution of the 
$1h_{11/2}$, $1g_{9/2}$ and $1g_{7/2}$ proton states, rather than by that of the 
$2d_{3/2}$, $2d_{5/2}$ or $3s_{1/2}$ proton states.
Let us mention 
also that in the nuclei considered there are more protons of the 
first type than those of the second type. Moreover,
the values $r_{ck}$, for the states of the first type, are
increasing functions of A for $A\ge 208$ while the respective values for 
the states of the second type almost remain constant with
the increasing of A in this region. One can effectively describe this type of 
behavior as a result of the repulsive interaction experienced by the 
$1h_{11/2}$, $1g_{9/2}$ and $1g_{7/2}$ protons which pushes 
them outside the central part of the nucleus and 
increases the value of $r_c$.
From Fig. 3 it is also seen that the behavior of the total value of 
$r_c$ for the nuclei in the interval $206\le A\le 208$
is strongly affected by the $r_{ck}$ value for $k=3s_{1/2}$.
In fact, a similar influence is also shown by the $2s_{1/2}$ state, 
in spite of its very large binding energy. In cases where the $\rho$ 
contribution to the ($S+V$) term in the central potential is reduced 
(cases $b$ and $d$ in Fig. 2), the contributions of the two $2s_{1/2}$ and
$3s_{1/2}$ states to $\Delta r^-_c=r_c(^{208}Pb)-r_c(^{206}Pb)$ 
are even larger. The two neutrons added to the $3p_{1/2}$ state
appear to be very effective to pull on the protons in these two states.

Table I and Fig. 2 also show 
that in the region $206 \le A \le 208$ it 
is easier to polarize protons by adding 
neutrons when protons are less bound (cases $b$ and $d$), in accordance 
with the reasons given in Ref. \cite{rein}. 
However, the behavior of $r_c$ in the region 
$208 \le A \le 210$ is the opposite. 
The $r_c$ behavior shows that the cases that produce
larger variation of $\Delta r^-_c$ are less effective contributing
to $\Delta r^+_c=r_c(^{210}Pb)-r_c(^{208}Pb)$. 
This is a reflection of the fact that it is more difficult 
to increase the $r_c$ value by adding extra neutrons when 
the radius is larger than in the case when it is smaller 
and when the neutrons previously added are more effective 
to increase $r_c$.

As explained above, although the $\rho$ meson contribution to the 
total energy is rather small in comparison with that of the $\sigma$ and 
$\omega$ mesons, it is just the $\rho$ meson that is mainly responsible 
in the RMFA for the kink structure. 
However, even without $\rho$ meson contribution ($g_\rho=0$)
there is a residual kink, case $d$ in Fig. 2. 
In this case, we have investigated
the role of the different terms entering
$V_{cent}$ and of
$V_{SO}$ in producing the residual kink.
To do that, we have artificially changed the contribution of these terms 
in a similar way as we did it for the different components of the $\rho$ meson, 
but 
readjusting now the strength of the ($S+V$) term, if necessary, to keep the
$r_c$ values close to the experimental ones for the $Pb$ isotopes 
studied,
(the same conditions are used for the three isotopes).
These studies reveal that the terms $\ve\frac{V}{M}$, 
$\Delta V_{cent}$ and $V_{SO}$
contribute in a slight and 
positive way to the kink structure, 
whereas the term $(S^2-V^2)/2M$ has an opposite effect 
that approximately cancels the effects mentioned above. 
We notice that taking about the 70\% of the 
strength of the ($S+V$) term entering Eq. (3)
as the total potential, the KE (obtained in the $Pb$ isotopes 
studied taking $g_\rho=0$) is mainly reproduced
as an effect of the self-consistent procedure. For example, if the $S$ 
and $V$ potentials are approached by the Woods-Saxon functions
with the radius  proportional to $A^{1/3}$ the KE disappears, 
although different diffuseness parameters for the potentials are used.

We have studied above the role 
that $V_{cent}$ and $V_{SO}$ play in generating and forming the structure 
of the kink.
Another crucial relativistic effect comes from the fact that the source 
term in the scalar field equation, the scalar density ($\rho_S$), 
is smaller than the actual baryon density ($\rho_B$). 
This difference of two densities is just the mechanism that brings 
about the saturation in the RMFA, 
and essentially distinguishes relativistic models from nonrelativistic ones.
Thus it would be interesting to check also the role of this 
mechanism in the KE.
One naive way would be to replace $\rho_S$ 
in the scalar field equation by $\rho_B$.
Since the nucleus would become unstable in this 
case, we add to $V_{cent}$ in the single-particle Schr\"odinger equation (1)
a repulsive density-dependent term formally identical to the term proportional to 
$t_3$ appearing in the corresponding equation for the Skyrme-Hartree-Fock 
model.
The coefficient $t_3$ is chosen to fit the $r_c(^{208}Pb)$.
Under these conditions we observe that in two cases, 
with and without the $\rho$ meson (cases a and d, respectively), the quantities 
$r^+_c$ and $r^-_c$ are reduced and so is the ratio 
${r^+_c}/{r^-_c}$.
Thus, the KE is reduced about 15\% in the case $a$ and 60\% in the case 
$d$. The residual KE without the $\rho$ meson almost disappears.

Our conclusions can be summarized in the following points:

1. The KE in Pb isotopes chain appears to be a general feature of the 
RMFA. It is observed in both linear and nonlinear models, being more strongly 
manifested in the models with smaller values of K.
We have found that the $\rho$ meson plays an essential role in forming 
the KE. In the case without the $\rho$ meson a small kink remains that 
can be attributed to relativistic as well as to self-consistent 
effects.

2. The KE mainly comes from the (partially) 
destructive interference of 
contributions of the $\rho$ meson to the ($S+V$) and $(S^2-V^2)/2M$
components of the single-particle potential. The ($S+V$) term leads to 
a KE of positive sign while the $(S^2-V^2)/2M$ 
component produces a KE of negative sign.
The total effect has a positive sign and corresponds to the 
trend observed empirically (see cases $b$ and $c$ in Fig. 2).

3. Excluding the $\rho$ contribution from the 
two components of $V_{cent}$ mentioned in the previous point, 
i.e., taking $x_1=x_2=0$, one gets a strong reduction of the KE.
This situation is quite similar to that in which the $\rho$ meson 
contribution is completely excluded ($g_\rho$=0, case $d$ in Fig. 2). 
The residual KE in this case is related to the self-consistent 
procedure. 
This effect completely disappears if one uses a Woods-Saxon potential in 
the Dirac equation rather than a self-consistent calculation.

4. The contribution of the $\rho$ 
meson to $V_{SO}$ has no influence on the KE. 

5. The relativistic effect that produces saturation in the 
RMFA provided by the difference between the scalar density 
and the baryon density, also contributes to a certain increase of the KE 
with respect to the nonrelativistic case in which saturation is produced by 
a repulsive density-dependent term as in the Skyrme-Hartree-Fock 
model.

In a forthcoming publication we shall investigate the problem 
of the KE in the relativistic Hartree-Fock framework.

One of the authors (L.N.S.) is very grateful to the University of Cantabria
and to the Technical University of Munich for hospitality.
This work has been supported in part by the DGESIC grant PB97-0360.

\newpage

\section*{Figure Captions} 

Fig. 1. The charge radii of $Pb$ isotopes obtained with the L 
\cite{horo81}, NL-SH \cite{shar93b} and NL3 \cite{lala97} 
models, normalized to the experimental value of the $^{208}Pb$ nucleus.
The empirical values (EXP) are taking from Ref. \cite{shar93b}. 

\bigskip

\noindent
Fig. 2. The same as in Fig. 1 for the NL-SH \cite{shar93b} model,
and the four cases considered in Table I:
$a$: $x_{1-5}=1$ (the exact model);
$b$: $x_1=0.1$, $x_{2-5}=1$;
$c$: $x_{1,3-5}=1$, $x_2=0$;
$d$: $x_{1-5}=0$ (without the $\rho$ meson contribution).

\bigskip

\noindent
Fig. 3. The single-particle r.m.s. radii for the proton states, as indicated, 
in the $Pb$ isotopes obtained with the NL-SH \cite{shar93b} model.

\newpage

\begin{table} [ht] 
\centering
\begin{tabular}{@{\extracolsep{2.0 mm}}lccccccc}
\hline
Model&$r_c$(fm)&$\Delta r^+_c$(fm)&
$\Delta r^-_c$(fm)&$\Delta r^+_c/\Delta r^-_c$&case\\ 
\hline 
L     & 5.455 & 0.016 & 0.010 & 1.59 & $a$\\
NL-SH & 5.502 & 0.020 & 0.010 & 1.93 & $a$\\
NL3   & 5.517 & 0.023 & 0.010 & 2.37 & $a$\\
\hline   
L     & 5.577 & 0.016 & 0.018 & 0.88 & $b$\\
NL-SH & 5.651 & 0.018 & 0.022 & 0.82 & $b$\\
NL3   & 5.718 & 0.019 & 0.022 & 0.85 & $b$\\
\hline 
L     & 5.421 & 0.016 & 0.008 & 1.99 & $c$\\
NL-SH & 5.477 & 0.020 & 0.008 & 2.46 & $c$\\
NL3   & 5.495 & 0.024 & 0.008 & 3.08 & $c$\\
\hline   
L     & 5.510 & 0.016 & 0.014 & 1.17 & $d$\\
NL-SH & 5.558 & 0.018 & 0.014 & 1.28 & $d$\\
NL3   & 5.568 & 0.021 & 0.013 & 1.58 & $d$\\
\hline   
Exp   & 5.503 & 0.019 & 0.011 & 1.73 &    \\
\hline   
\end{tabular}  
\caption {The charge radius ($r_c$), 
$\Delta r^+_c=r_c(^{210}Pb)-r_c(^{208}Pb)$, 
$\Delta r^-_c=r_c(^{208}Pb)-r_c(^{206}Pb)$ and 
the ratio $\Delta r^+_c/\Delta r^-_c$  for the linear 
model L [9] and the nonlinear models NL-SH [10] and NL3 [11], for the cases:
$a$: $x_{1-5}=1$;
$b$: $x_1=0.2$ (L), 0.1 (NL-SH), 0 (NL3), $x_{2-5}=1$;
$c$: $x_{1,3-5}=1$, $x_2=0$;
$d$: $x_{1-5}=0$. The last line contains the experimental results [1].}
\label{table I} 
\end{table}
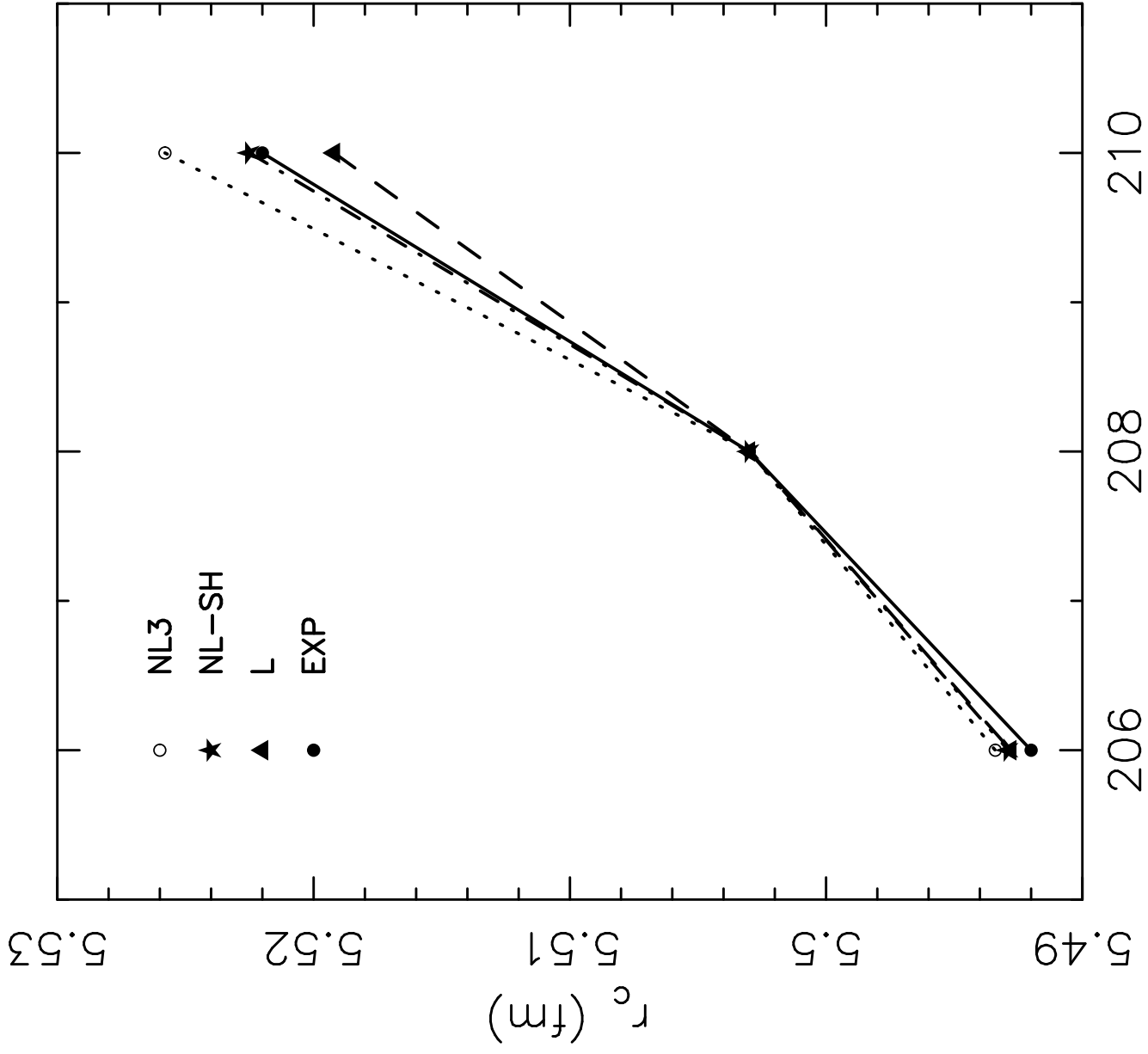
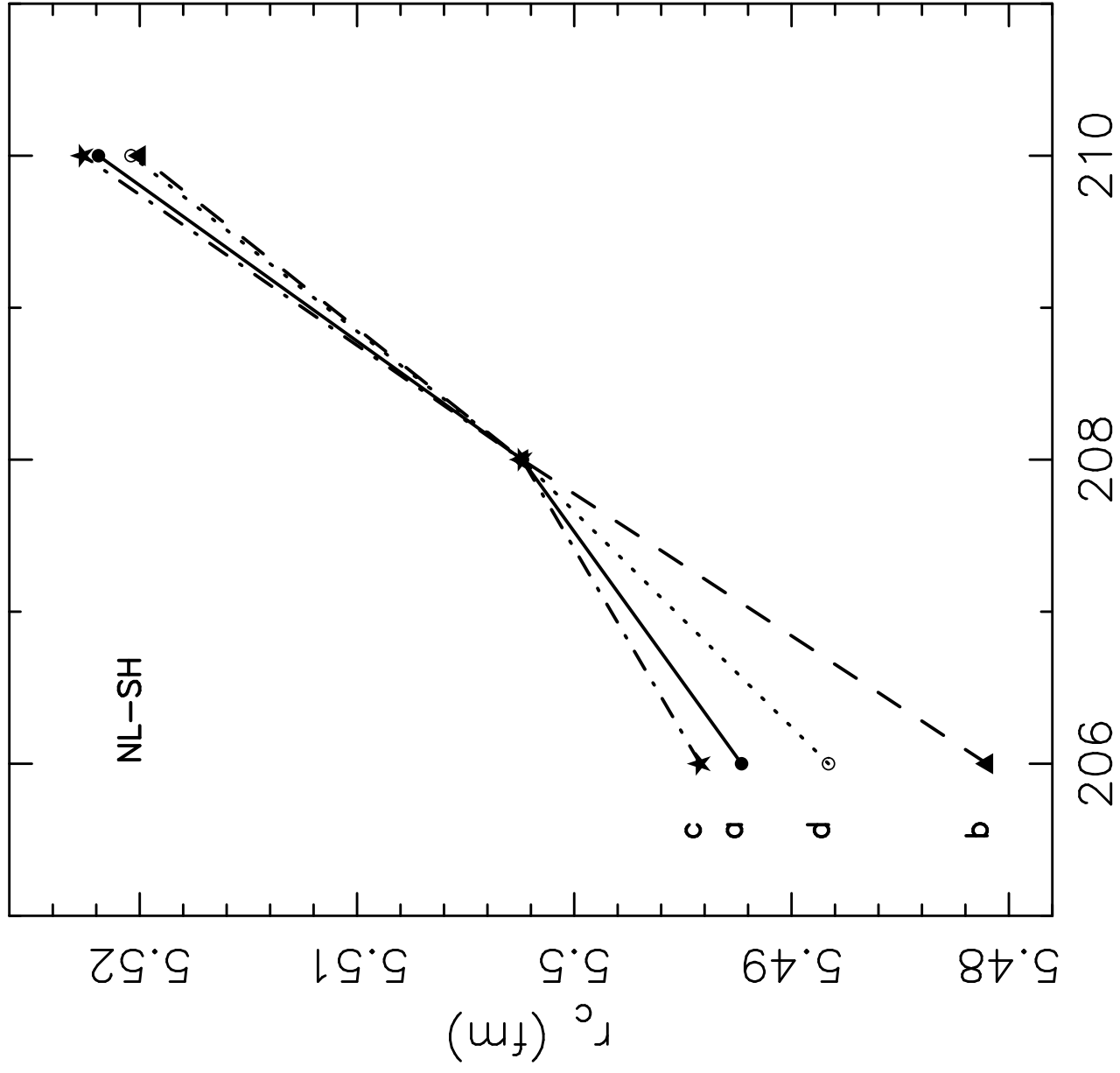
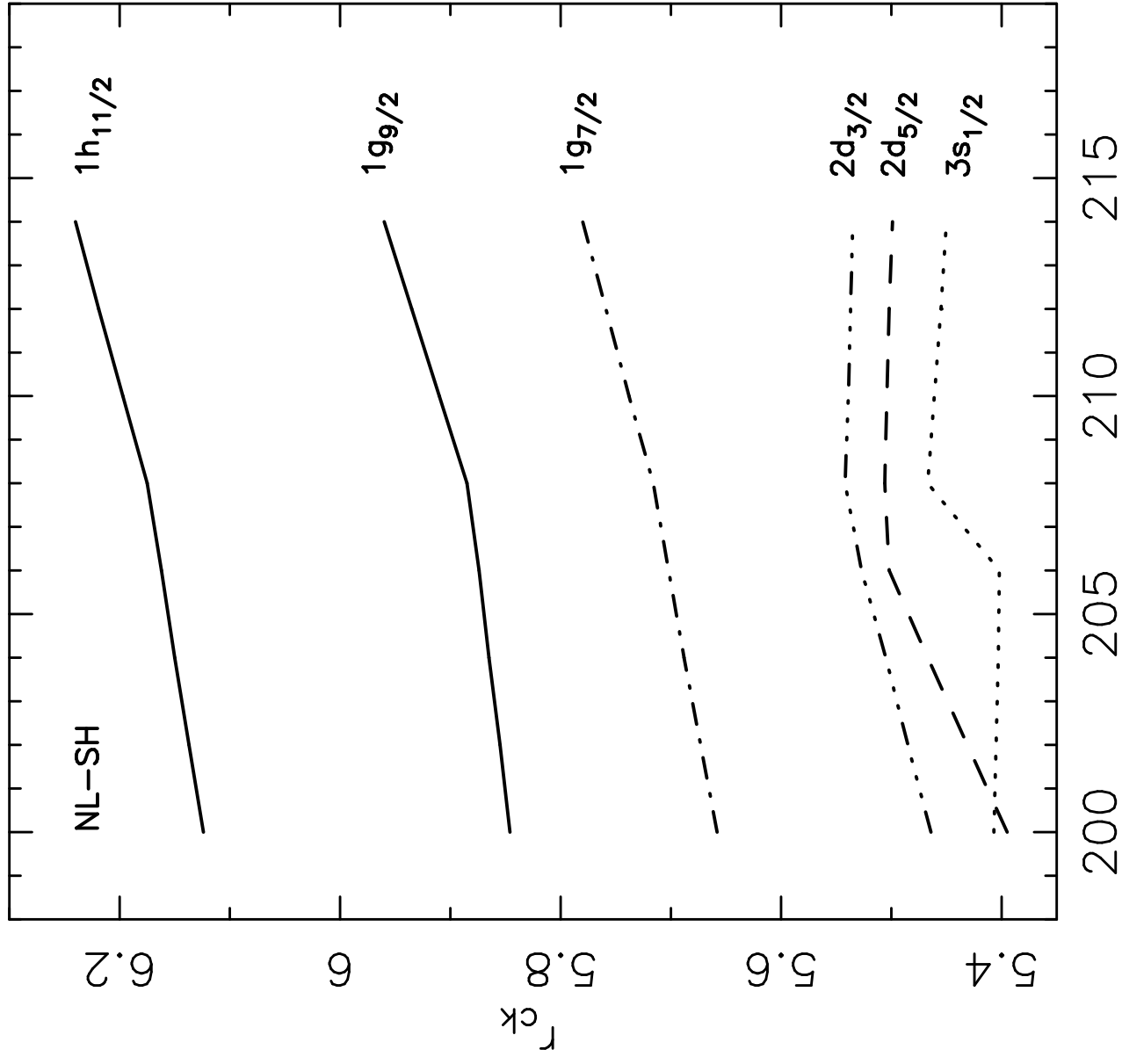
\end{document}